\documentclass[11pt,tightenlines]{revtex4}

\usepackage{amsmath}
\usepackage{amssymb}
\usepackage{bbm}
\usepackage{mathtools}
\usepackage[normalem]{ulem}
\usepackage{dsfont}
\usepackage{mathrsfs}
\usepackage{cancel}

\usepackage{braket}

\usepackage[utf8]{inputenc}
\usepackage[T1]{fontenc}

\def\eq{\text{eq}}

\def\la{\langle}
\def\ra{\rangle}

\def\ii{\mathrm{i}}

\def\B#1{\!\left(#1\right)}
\def\BB#1{\!\left[#1\right]}
\def\BBB#1{\!\left|#1\right|}

\def\be{\begin{equation}}
\def\ee{\end{equation}}

\def\bee{\begin{equation*}}
\def\eee{\end{equation*}}

\def\ba{\begin{equation}\begin{aligned}}
\def\ea{\end{aligned}\end{equation}}

\def\pgs{p_{GS}}

\def\pref{0.575}

\def\for{\ \text{for} \ }

\makeatletter
\newcommand{\pushright}[1]{\ifmeasuring@#1\else\omit\hfill$\displaystyle#1$\fi\ignorespaces}
\newcommand{\pushleft}[1]{\ifmeasuring@#1\else\omit$\displaystyle#1$\hfill\fi\ignorespaces}
\makeatother

\usepackage[ddmmyyyy]{datetime}
\makeatletter
\def\Dated@name{}
\makeatother

\begin{document}

\title{One-half of the Kibble-Zurek quench followed by free evolution}
\author{Micha{\l}  Bia\l ończyk and Bogdan Damski}
\affiliation{Jagiellonian University, Institute of Physics, {\L}ojasiewicza 11, 30-348 Krak\'ow, Poland}
\begin{abstract}
We drive the one-dimensional quantum Ising chain in the transverse field from the 
paramagnetic phase to the critical point and study its free evolution there.
We analyze excitation of such a system at the critical point and 
dynamics of its transverse magnetization and  Loschmidt echo during free
evolution. We discuss how  the system size and  quench-induced scaling
relations from the Kibble-Zurek theory of non-equilibrium phase transitions are encoded in 
quasi-periodic time evolution  of the transverse magnetization and Loschmidt echo.    
\end{abstract}
\maketitle

\section{Introduction}
\label{Introduction}
Suppose we prepare a system, which possesses  a quantum phase transition, in
the ground state of a gapped phase far away from the critical point. Then, we drive it through a gradual
change of some parameter of its Hamiltonian   to the critical 
point, where changes of the Hamiltonian stop and the system is let to undergo free evolution. 
The question we would like to address here is how  such  free evolution
reflects the universal  non-equilibrium excitation of the system and the fact
that it takes place at the critical point.

The above quench protocol can be seen as combination of  continuous  and sudden 
quench protocols that have been  intensively studied lately. In the former,
the system is driven through a smooth  change of some  parameter across the critical point. 
Its
non-equilibrium final state is then typically analyzed  
far away from the critical point  in the new phase 
\cite{JacekAdv2010,Dutta2015}.
In the latter, the instantaneous change of some parameter is 
imposed on the system and subsequent non-equilibrium free evolution  
is  analyzed \cite{JacekAdv2010,PolkovnikovRMP2011}. In our case, we focus on  free evolution, just as in the
sudden quench, but take as the initial state the non-equilibrium state of the
system resulting from  continuous driving  to the critical point.
That such a quench protocol can lead to interesting results can be seen in the
following way. 

Prepare finite but large-enough system (\ref{Ngg}) in the ground state, assume that its 
evolution towards the critical point is initially 
adiabatic,  and choose linear in time change of the external field driving the
transition
\be
g(t)=g_c-\frac{t}{\tau_Q},
\label{godt}
\ee
where  $g_c$ is the critical point and $\tau_Q$ is the quench time
inversely proportional to the driving rate. Evolution of the system cannot be adiabatic all the way to
the critical point because  the gap in the excitation spectrum  
becomes too  small near the critical point. The crossover from 
adiabatic to non-adiabatic dynamics  is then  
studied through the quantum version \cite{BDPRL2005,DornerPRL2005,JacekPRL2005} of the Kibble-Zurek theory of
non-equilibrium phase transitions \cite{KibbleRev,ZurekRev,delCampoReva,delCampoRevb}. 

One notices in this approach that there are two time scales determining system's dynamics.
The first one is proportional to the inverse of the instantaneous energy gap $\Delta$ between the ground
state and the first excited state. It can be seen as the
reaction time of the system: the larger the gap is, the quicker the system  adjusts to  driving. 
The second time scale quantifies how fast the system is
driven and so it is inversely  proportional to rate at which the gap changes, 
$\Delta/\frac{d}{dt}\Delta$. The crossover between the adiabatic regime
and the non-equilibrium one takes place when these two time scales become
comparable
\be
\frac{\hbar}{\Delta} \sim \frac{\Delta}{\frac{d}{dt}\Delta}.
\ee
Such an equation can be solved by combining (\ref{godt}) with the standard scaling relation $\Delta\sim|g_c-g|^{z\nu}$, 
where $z$ and $\nu$ are the dynamical and correlation-length exponents,
respectively. The solution gives  scaling of the  distance $\hat g$ from the critical point,
where  the crossover takes place 
\be
\hat g \sim\tau_Q^{-1/(1+z\nu)}.
\label{ghat}
\ee

Using this result, one can also obtain scaling of the time that is left to
reaching the critical point from the moment when the system enters the
crossover regime
\be
\hat t= \hat g \tau_Q\sim\tau_Q^{z\nu/(1+z\nu)}.
\label{that}
\ee

Besides the  characteristic field and time scales, $\hat g$ and $\hat t$, respectively,  the Kibble-Zurek theory
implies existence of the characteristic quench-induced  length scale
\be
\hat\xi=\xi(\hat g)\sim\tau_Q^{\nu/(1+z\nu)},
\label{xihat}
\ee
where $\xi$ is the equilibrium correlation length. Finite-size effects in
the Kibble-Zurek quench  (\ref{godt}) are 
 negligible when the system size $N$ satisfies
\be
N\gg\hat\xi.
\label{Ngg}
\ee

Different observables, computed in the non-equilibrium  state of the system
after  the Kibble-Zurek quench,  should scale non-trivially with the
quench rate akin to (\ref{ghat})--(\ref{xihat}). 
It is now interesting to ask how they would behave in the time domain during
free evolution at the critical point?

The only quench-independent characteristic time scale that we have at the critical point, which  is
relevant for slowly driven systems being of interest here, is the inverse of the typical 
energy gap between low-lying energy levels.
Taking the standard dispersion relation at the critical point, 
\be
\omega(k)\sim k^z,
\ee
and noting that $k\sim 1/N$, one finds the characteristic time scale 
\be
\tau_c\sim N^z,
\label{tauc}
\ee
which is divergent in the thermodynamic limit.
Before moving on, we note that  (\ref{Ngg}) is equivalent to the
condition that the gap between the ground state and the first excited state at
the distance $\hat g$ from the critical point, which is proportional to 
$\hat g^{z\nu}\sim1/\hat t$, is much larger than the gap at the 
critical point that is proportional to $1/\tau_c$. This leads to 
\be
\tau_c\gg\hat t,
\ee
which  is the same as (\ref{Ngg}).

The goal of this work is to discuss how  Kibble-Zurek scaling relations (\ref{ghat})--(\ref{xihat}) and 
system-size dependent time scale (\ref{tauc}) are encoded in  free evolution of the one-dimensional 
quantum Ising model in the transverse field that was driven to the critical point. 
This system is a  standard testbed for various studies of non-equilibrium dynamics. Different aspects of 
its continuous quenches  were researched  in numerous references, see e.g.
\cite{JacekPRL2005,DornerPRL2005,PolkovnikovPRB2005,RalfPRA2007,PolkovnikovPRL2008,SenPRB2009,SantoroPRB2009,JacekPRB2009,SenPRA2009,ArnabPRB2010,KolodrubetzPRL2012,FrancuzPRB2016,SantoroJstat2015,PuskarovSciP2016,ApollaroSciRep2017}. 
Significant efforts were  also devoted to investigations of sudden quenches in
this model, see e.g. 
\cite{SenguptaPRA2004,CalabreseJstat2012a,CalabreseJstat2012b,ArnabPRB2012,VedralPRL2012,EsslerPRL2012,HeylPRL2013,ArnabSciRep2015}.

Free dynamics of a quantum system driven to the critical point was
theoretically studied in
 \cite{BDNJP2008}, where the system of interest was a weakly-correlated spin-1 Bose-Einstein condensate. 
Its dynamics   was modeled by the mean-field theory approximating
system's  behavior in the thermodynamic  limit. It
was found that non-trivial oscillations of magnetization take place at the
critical point during free  evolution following the Kibble-Zurek quench. 
Their non-triviality  followed from 
the fact that their amplitude was inversely proportional to  their period, which was proportional to $\hat t$.  
It is of interest in this work to find out whether  comparably-interesting dynamics can
be found in  a strongly-correlated  many-body system such as the quantum Ising chain.

This paper is organized as follows. Basic information about the quantum Ising
model is provided in Sec. \ref{Model_sec}. A continuous quench of this system 
towards the critical point is described in Sec. \ref{Dynamics_towards}. 
Dynamics of the transverse magnetization and
Loschmidt echo during free evolution at the critical point is studied in 
Secs. \ref{Dynamics_sec} and
\ref{Loschmidt_sec}, respectively. Brief summary of our results is available in  Sec. \ref{Summary_sec}.

\section{Model}
\label{Model_sec}
We study the one-dimensional quantum  Ising model in the transverse field  described by the following Hamiltonian
\be
\hat H(g) = -\sum_{i=1}^{N}\B{\sigma^x_i\sigma^x_{i+1} + g \sigma^z_i},
\label{HIsing}
\ee
where $g$ is the external magnetic field and   $N$ is the number of spins. We impose periodic boundary conditions 
on the system and take even $N$. 

This model undergoes  a quantum phase transition and it is exactly solvable,
which greatly facilitates its study. Its basic equilibrium properties  were described in the following seminal
references  \cite{Lieb1961,Pfeuty}.  For the purpose of this work, we need to
state that there is the critical point $g_c=1$ separating the ferromagnetic phase ($0<g<1$)
from the paramagnetic phase ($g>1$). The critical exponents of this model
are $z=\nu=1$. This means that 
\be
\label{wczasy}
\hat t\sim \sqrt{\tau_Q}, \  \tau_c\sim N,
\ee
\be
\hat\xi\sim\sqrt{\tau_Q}.
\label{hatxi}
\ee
We will consider below  system sizes
\be
N\gg\sqrt{\tau_Q}
\label{Nggsqrt}
\ee
to satisfy condition (\ref{Ngg}).

The quench protocol, which  we have  proposed in Sec. \ref{Introduction}, reads 
\be
g(t)=
\left\{
\begin{array}{ll}
1-t/\tau_Q & \for  t \le 0 \\
1 & \for t> 0 
\end{array}
\right..
\label{qprotocol}
\ee
Evolution starts from the ground state of Hamiltonian (\ref{HIsing})  deeply in the paramagnetic  phase and 
the system evolves  towards the critical point arriving there in an excited state at $t=0$. 
 The degree of excitation is controlled by  the quench time $\tau_Q$ 
 (larger $\tau_Q$ leads to more adiabatic  evolution).
Once the critical point is reached, the magnetic field is no longer changed so that the
system undergoes  free evolution there.

\section{Dynamics towards the critical point}
\label{Dynamics_towards}
Time evolution of the quantum Ising  model  can be most conveniently
described by mapping spins onto  non-interacting fermions through the Jordan-Wigner
transformation \cite{JacekPRL2005}. Such a transformation reads
\be
\sigma^z_i=1-2\hat c_i^\dag\hat c_i, \ \ \sigma^x_i=(\hat c_i+\hat c_i^\dag)\prod_{j<i}\B{1-2\hat c_j^\dag\hat c_j},
\ee
where $\hat c_i$ are fermionic operators.
One then notes that the Hamiltonian  commutes with the parity operator 
\be
\prod_{i=1}^N\sigma^z_i,
\ee
whose eigenvalues are $\pm1$. This has two consequences. First,  eigenvalues of $\hat H$ can be
assigned definite parity, positive or negative. 
Second, if the system is initially prepared in the state  with definite parity, then 
during time evolution only states with such parity are populated.
We start evolution from the ground state, which for even-sized systems that we consider, 
has  positive parity \cite{BDJPA2014}.
Therefore, we focus below on the dynamics in the positive-parity subspace of the Hilbert
space, where anti-periodic boundary conditions are imposed on fermionic
operators: $\hat c_{N+1}=-\hat c_1$ \cite{BDJPA2014}.

One then goes to the  momentum space through the substitution 
\begin{align}
&\hat c_j= \frac{\exp(-\ii\pi/4)}{\sqrt{N}}\sum_{K=\pm k} \hat c_K\exp(\ii K
j),\\
&k=\frac{\pi}{N},\frac{3\pi}{N},\cdots,\pi-\frac{\pi}{N},
\label{kk}
\end{align}
where $\hat c_{\pm k}$ annihilates  a quasi-particle with quasi-momentum $\pm k$.
After these transformations,  one obtains
\be
\hat H = 
2\sum_k  (\hat c_k^\dag \hat c_k -\hat c_{-k} \hat c_{-k}^\dag) [g-\cos(k)]
+  (\hat c_k^\dag\hat c_{-k}^\dag + \hat c_{-k}\hat c_{k})\sin(k),
\ee
which can be diagonalized through the Bogolubov transformation. The ground state of this Hamiltonian is
\begin{align}
& |g\ra=\prod_{k} 
 \B{u_k^\text{eq} - v_k^\text{eq}\hat c_k^\dag\hat c_{-k}^\dag}|{\rm vac}\rangle, \\
 \label{uveq}
&u_k^\text{eq}=\cos\B{\frac{\theta_k}{2}}, \ v_k^\text{eq}=\sin\B{\frac{\theta_k}{2}}, \\
&\sin \theta_k = \frac{\sin(k)}{\sqrt{g^2-2g\cos(k)+1}}, \ \ \cos\theta_k= \frac{g-\cos(k)}{\sqrt{g^2-2g\cos(k)+1}},
\end{align}
where 
 $u_k^\eq$ and $v_k^\eq$ are the equilibrium Bogolubov modes and 
$|{\rm vac}\rangle$ is annihilated by all $\hat c_{\pm k}$ operators.

Solving the time-dependent Schr\"odinger equation, $\ii\frac{d}{dt}|\psi(t)\rangle=\hat H[g(t)]|\psi(t)\rangle$,  one finds
that  dynamics of the
$N$ spin system splits into  dynamics of $N/2$ uncoupled Landau-Zener
systems. The wave-function can be then written as   \cite{JacekPRL2005}
\be
|\psi(t)\rangle =\prod_{k} \B{u_k(t) - v_k(t)\hat c_k^\dag\hat c_{-k}^\dag}|{\rm vac}\rangle,
\label{wavef}
\ee
where $u_k$ and $v_k$ are the time-dependent  Bogolubov modes evolving  according to 
\be
\ii\frac{d}{dt}
\left(
\begin{array}{c}
v_k \\
u_k
\end{array}
\right)
=h_k
\left(
\begin{array}{c}
v_k \\
u_k
\end{array}
\right), \ h_k=
2\left(
\begin{array}{cc}
g(t)-\cos(k) & -\sin(k)\\
-\sin(k)  & \cos(k)-g(t)
\end{array}
\right).
\label{lz}
\ee

Evolution starts at $t=-\infty$ from the state, where all spins are
perfectly aligned with the magnetic field. In the  non-interacting 
fermions formalism  this translates into 
the following initial conditions  
\be
|u_k(t=-\infty)|=1, \ v_k(t=-\infty)=0.
\ee

This means that all two-level  systems (\ref{lz})  start evolution from the 
ground state of the Hamiltonian $h_k$ and  results in the following exact solution 
\begin{align}
\label{vkodt}
&v_k(t) = \sqrt{\tau_Q}\sin(k) e^{-\pi\tau_Q\sin^2(k)/4}D_{-n-1}[\ii z(t)],\\
\label{ukodt}
&u_k(t) = e^{-\pi\tau_Q\sin^2(k)/4}e^{-\ii\pi/4}\B{(1+n) D_{-n-2}[\ii z(t)] +\ii z(t) D_{-n-1}[\ii z(t)]},\\
& \ii z(t) = 2\sqrt{\tau_Q} e^{\ii\pi/4}[g(t)-\cos(k)],  \ n = \ii\tau_Q\sin^2(k),
\end{align}
where $D_m$ is the Weber function \cite{Watson}. 
We obtained such a solution 
 by mapping  (\ref{lz}) into the Landau-Zener 
form \cite{JacekPRL2005} 
\be
\begin{aligned}
\label{lz1}
&\ii\frac{d}{dt'}
\left(
\begin{array}{c}
u_{k} \\
v_{k}
\end{array}
\right)
= 
\frac{1}{2}\left(
\begin{array}{lr}
t'/\tau_Q' & 1 \\
1  & -t'/\tau_Q' 
\end{array}
\right)
\left(
\begin{array}{c}
u_{k} \\
v_{k}
\end{array}
\right),\\
&t'=4\tau_Q\sin(k)[g(t)-\cos(k)],  \ \tau_Q'=4\tau_Q\sin^2(k),
\end{aligned}
\ee
and then adopting to the current problem the steps from Appendix B of
\cite{BDPRA2006}. Eqs. (\ref{vkodt}) and (\ref{ukodt}) can be
used for deriving analytical approximations. They are, however, impractical
for getting exact results, which we obtain by direct numerical integration  of   (\ref{lz}).

Having these results, we can  quantify the level of excitation at the critical point
to better characterize the initial state for  subsequent free evolution
that we describe in Secs. \ref{Dynamics_sec} and \ref{Loschmidt_sec}.  We compute the
probability of finding the Ising chain in the ground state at $t=0$
\be
\pgs=\prod_k (1-p_k)=\exp\B{\sum_k \ln(1- p_k)},
\label{pgs}
\ee
where $p_k$ is the probability of finding two-level system (\ref{lz})
in the {\it excited} state at $t=0$. 
 We could, of course, write
(\ref{pgs}) as the product of probabilities of finding two-level systems
(\ref{lz}) in the ground state, but such a representation is inconvenient given 
analytical approximations that we will use--see discussion around
(\ref{pgsPk}) and (\ref{Pk}).

A simple calculation then shows that 
\begin{align}
\label{pk}
&p_k=\BBB{v_{ck}^\eq\, u_k(0)-u_{ck}^\text{eq}\,v_k(0)}^2,\\
\label{vkeq}
&v_{ck}^\eq=\cos(k/4+\pi/4),\\
\label{ukeq}
&u_{ck}^\eq=\sin(k/4+\pi/4),
\end{align}
where $v_{ck}^\eq$ and $u_{ck}^\eq$ are the equilibrium  Bogolubov modes (\ref{uveq}) 
computed at the  critical point.

\begin{figure}[t]
\includegraphics[width=\pref\columnwidth,clip=true]{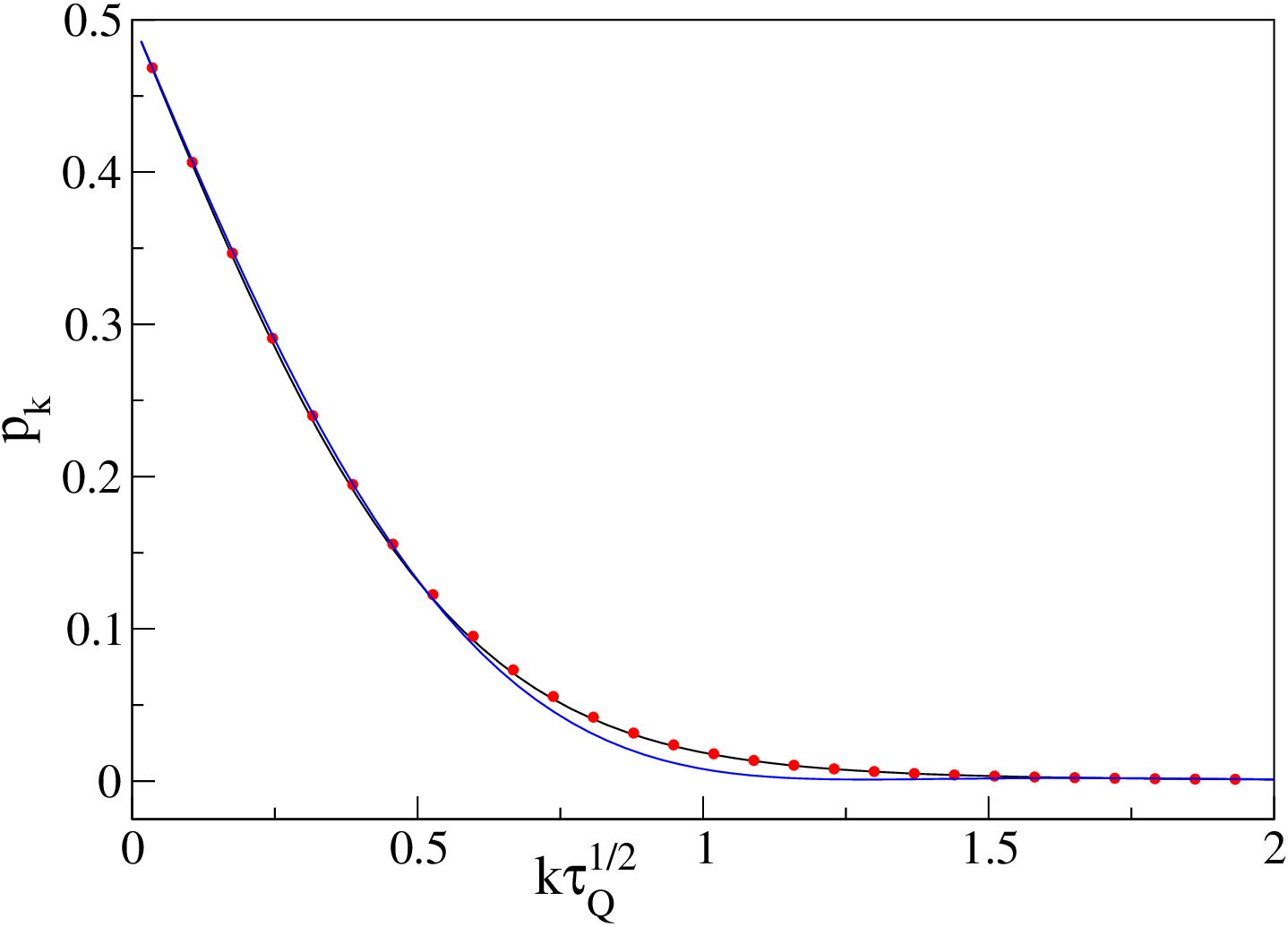}
\caption{Probability of finding  two-level system (\ref{lz}) in the
ground state at $t=0$.  The black line shows numerics for  $\tau_Q=100$ 
(data points are connected by a line). Red dots show numerics  for $\tau_Q=500$. 
The  blue line shows the approximate result obtained after putting
 (\ref{approx1})--(\ref{approx3}) into  (\ref{pk}). The system size $N=2000$.
}
\label{pk_fig}
\end{figure}

The basic expectation from the KZ theory is that small $k$ modes are excited
during evolution towards the critical point, i.e., 
\be
k\lesssim 1/\hat\xi\sim 1/\sqrt{\tau_Q}.
\label{khat}
\ee
As a result, we expect that for such momenta  $p_k$ is a function of
$k\sqrt{\tau_Q}$. This expectation cannot be exact as the argument of Weber
functions at the critical point is 
\be
iz(0) \sim \sqrt{\tau_Q}k^2=(k\sqrt{\tau_Q})^2/\sqrt{\tau_Q}
\label{iztc}
\ee
for small momenta. Nonetheless, such dependence is suppressed for large-enough $\tau_Q$. 
Thus, the following approximations, when put into (\ref{pk}), lead to fairly
accurate results
\begin{align}
\label{approx1}
&v_k (0) \approx \sqrt{\tau_Q} k \sqrt{\frac{\pi}{2}}e^{-\pi \tau_Q k^2/4},\\
\label{approx2}
&u_k (0) \approx e^{-\pi \tau_Q k^2/4} e^{-\ii \pi/4} \BB{1+\frac{\ii}{2}  \tau_Q k^2 (\gamma+\ln 2)},\\
\label{approx3}
&v_{ck}^\eq\approx u_{ck}^\eq  \approx \frac{1}{\sqrt{2}},
\end{align}
where $\gamma=0.5772\cdots$.
The scaling properties of $p_k$ are illustrated in Fig. \ref{pk_fig}. It is
worth to note that we expand in a series in $k$ the argument of  $e^{-\pi\tau_Q\sin^2(k)/4}$ instead of the whole 
exponential function. This  amounts to 
introducing the Gaussian cut-off function,
\be
e^{-\pi \tau_Q k^2/4},
\label{Gcutoff}
\ee
which makes  non-equilibrium
approximations (\ref{approx1}) and (\ref{approx2}) vanish for the momenta 
\be
k\gg 1/\hat\xi\sim1/\sqrt{\tau_Q}
\label{khatBis}
\ee
for which  equilibrium solutions 
(\ref{vkeq}) and (\ref{ukeq}) well  approximate 
exact non-equilibrium solutions (\ref{vkodt}) and (\ref{ukodt}), respectively. Such a
procedure allows us to separate the non-equilibrium effects from the
equilibrium ones, which   simplifies the following computations. It was first employed in  \cite{JacekPRL2005}.

We can now explain why it is convenient to use the  expression for $\pgs$ in the
form given by (\ref{pgs}).  Suppose, we write 
\be
\pgs=\prod_k P_k,
\label{pgsPk}
\ee
where $P_k$ is the probability of finding the two level system (\ref{lz}) in
the ground state. 
A simple calculation then yields 
\be
P_k= \BBB{v_{ck}^\eq\,v^*_k(0)+u_{ck}^\text{eq}\,u^*_k(0)}^2.
\label{Pk}
\ee
Putting approximations
(\ref{approx1}) and (\ref{approx2}) into (\ref{pk}) and (\ref{Pk}), we get 
for large-momenta, such as (\ref{khatBis}), that   $p_k,P_k\to0$.  In this limit, however,
we should be getting  
$p_k\to0$ and $P_k\to1$ because the modes with large-enough
momenta are only marginally  excited  during our time evolutions. 
As a result,  it is convenient to use expression (\ref{pgs}) instead of
(\ref{pgsPk}) to compute $\pgs$ because in the former case we can use 
approximate expressions for all momenta without introducing significant errors, 
while in the latter case 
 restriction to momenta satisfying (\ref{khat}) has to be enforced in the
product over $k$.

\begin{figure}[t]
\includegraphics[width=\pref\columnwidth,clip=true]{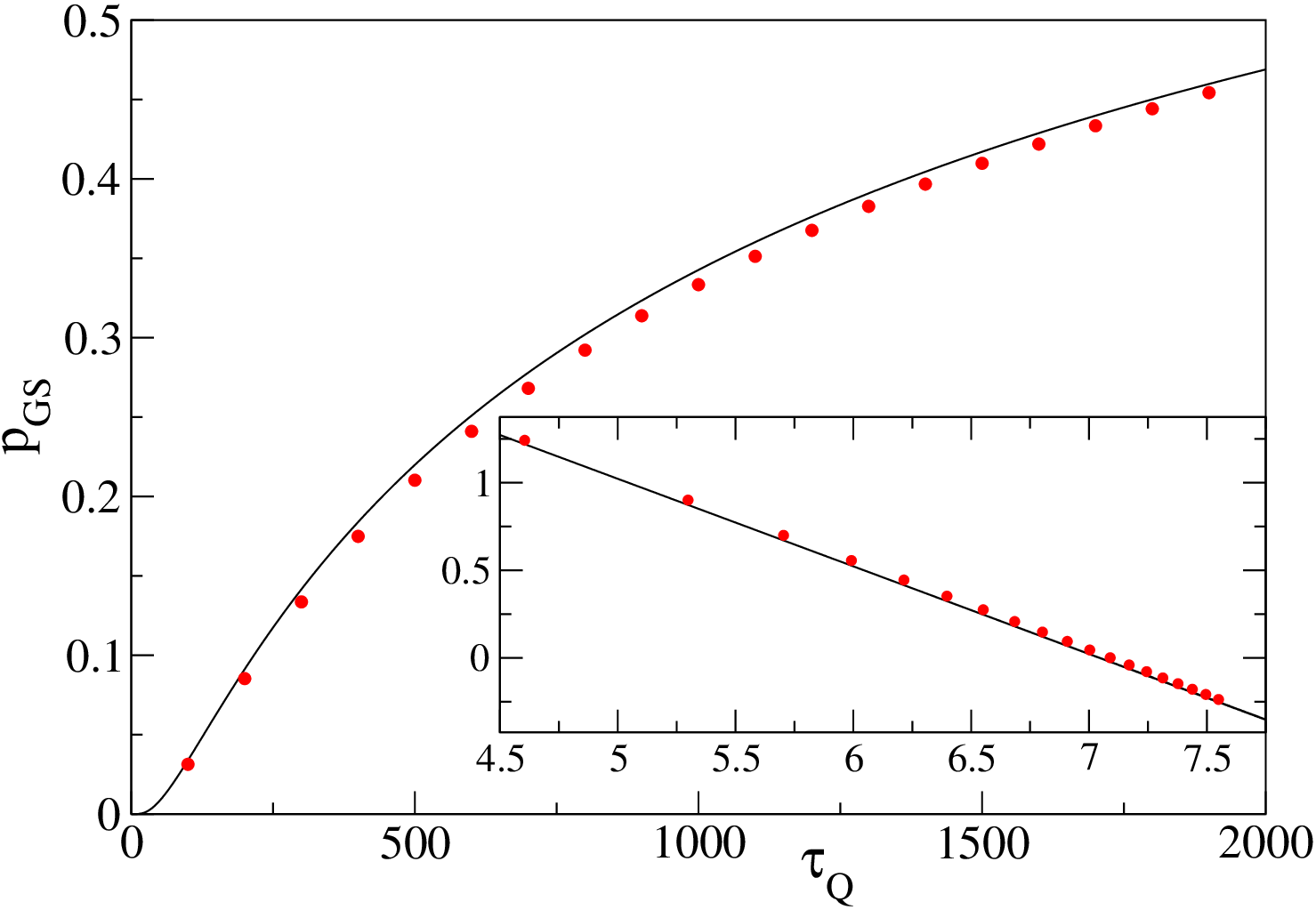}
\caption{Probability of finding the Ising chain   in the ground state at $t=0$.
Solid black line shows analytical approximation (\ref{exp_pgs_all}), while red dots provide numerics. 
The inset shows $\ln[-\ln(\pgs)]$ vs. $\ln(\tau_Q)$.
The system size $N=1000$.}
\label{pgs_fig}
\end{figure}

To further simplify  calculations, we  approximate for $N\gg1$ the sum in  (\ref{pgs}) by the integral 
\be
\sum_k \to \frac{N}{2\pi}\int_0^\pi dk,
\label{X}
\ee
which leads to 
\be
\pgs=\exp\B{\frac{N}{2\pi}\int_0^\pi dk \ln(1-p_k)}.
\ee
Putting  (\ref{approx1})--(\ref{approx3}) into  (\ref{pk}) and taking  $\tau_Q\gg1$, we can rewrite it  as
\begin{subequations}
\begin{align}
\label{exp_pgs}
&\pgs\approx\exp\B{-C\frac{N}{\sqrt{\tau_Q}}},\\
\label{CC}
&C =  - \frac{1}{2\pi}\int_0^{\infty} dx \ln \left ( 1 - \frac{1}{8} e^{-\pi x^2/2} [(x \sqrt{\pi}-2)^2+
(x \sqrt{\pi} - x^2 (\gamma + \ln 2))^2] \right )\approx0.034.
\end{align}
\label{exp_pgs_all}%
\end{subequations}
Comparison between (\ref{exp_pgs_all})  and numerics is presented in Fig.
\ref{pgs_fig}, where quite good agreement is found.

Result (\ref{exp_pgs}) can be also obtained through the adiabatic--impulse
approximation (AIA) introduced in the quantum context in  \cite{BDPRL2005,BDPRA2006}. 
 We will adopt this approximation to the problem studied in this paper.
In the spirit of AIA, we split evolution towards the critical point  into adiabatic and impulse
stages. The adiabatic stage takes place when the system is far away from the critical
point, which  happens when $g(t)-1>\hat g$.
AIA assumes  that populations of instantaneous eigenstates of the Hamiltonian 
do not  change when this condition is satisfied. The impulse 
stage takes place near   the
critical point, i.e. when  $g(t)-1<\hat g$. AIA assumes that the state of
the system does not change there. Combining these two assumptions with the fact that our evolution starts 
from the ground state, we see that 
within the AIA the system arrives at the critical point in the ground state
corresponding to the external field $g=1+\hat g$. Denoting  by 
$|g\ra$ the  ground state of $\hat H(g)$ (\ref{HIsing}), we find
\be
\pgs=|\la 1|1+\hat g\ra|^2
\label{aia1}
\ee
according to AIA. This equation expresses the non-equilibrium quantity, $\pgs$,
in terms of the equilibrium quantity, the overlap between ground states. Such
an overlap is known as fidelity and it plays an important role in the studies of 
equilibrium quantum phase transitions \cite{ZanardiPRE2006,GuReview}. It has been 
used in the context of dynamical phase transitions as well (see 
\cite{MarekBodzioPRA,BDproceedings} for results relevant to this work).
When  condition
(\ref{Ngg}) is satisfied, the following approximation can be derived
\cite{BDPRL2011}
\be
|\la 1|1+\hat g\ra|^2\approx\exp\B{-N\frac{\pi-2}{4\pi}\hat g}.
\label{aia2}
\ee
Combining (\ref{aia1}) and (\ref{aia2}) and taking $\hat g\sim 1/\sqrt{\tau_Q}$,
we get (\ref{exp_pgs}) with some unknown constant $C$, whose determination is
beyond AIA.

\begin{figure}[t]
\includegraphics[width=\pref\columnwidth,clip=true]{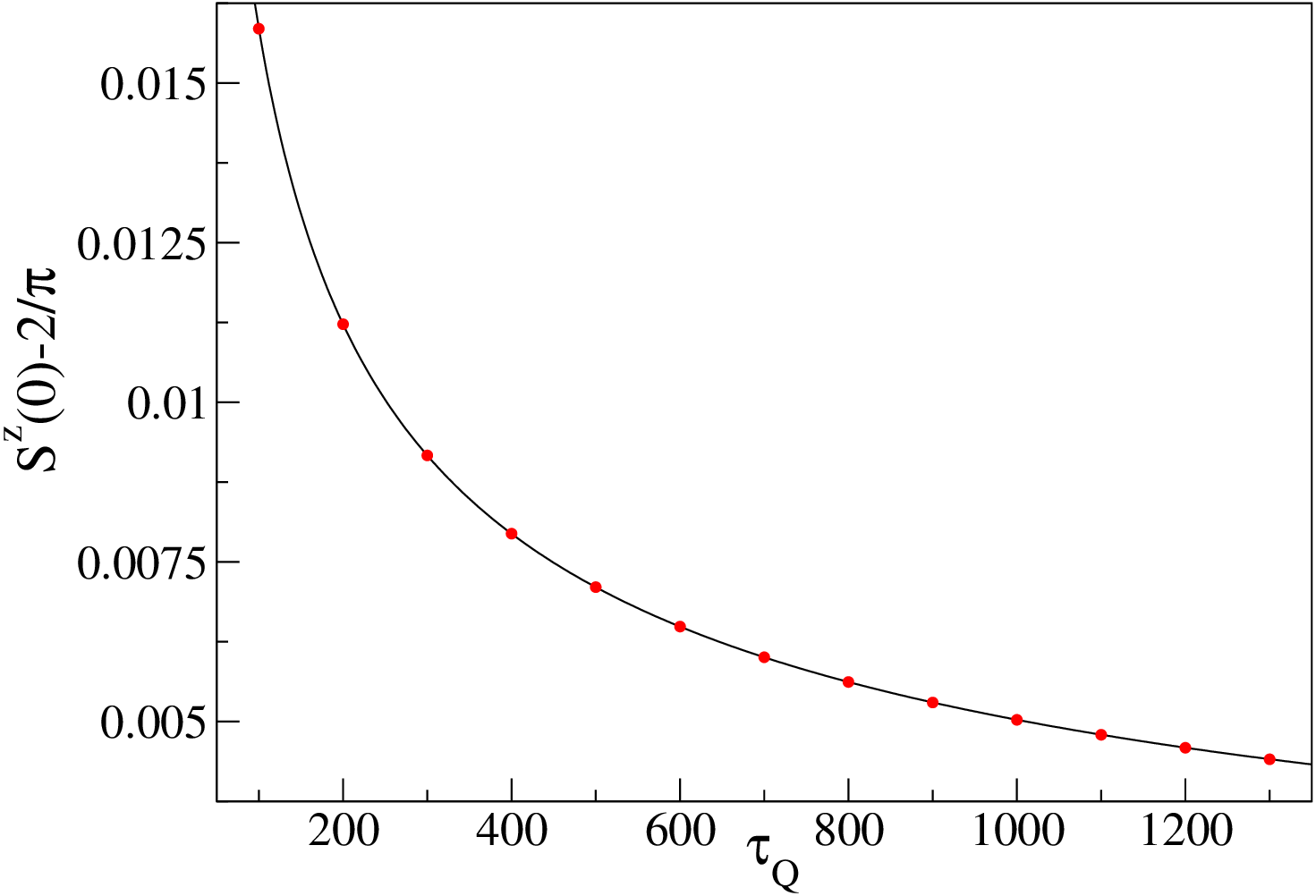}
\caption{Non-equilibrium transverse magnetization at the critical point less the equilibrium value of this 
observable. We approximate the latter by $2/\pi$, which is  the thermodynamic-limit
result differing negligibly from the one that can be numerically obtained in  the $N=1000$ system studied here. 
The black line comes
from the fit $\ln(S^z(0)-2/\pi)=-1.8465(5) - 0.4989(1)\ln(\tau_Q)$.
We report everywhere in this paper one standard error in the brackets next to
the fitted coefficients. 
}
\label{szcrit_fig}
\end{figure}

Finally, anticipating discussion in Sec. \ref{Dynamics_sec}, we introduce  
the transverse magnetization
\be
S^z(t)=\la\psi(t)|\sigma^z_i|\psi(t)\ra=1-\frac{4}{N}\sum_k |v_k(t)|^2.
\label{Sz}
\ee
We are interested now  in the value of this observable at time $t=0$, i.e., 
when the system  arrives at the critical point. 
In the limit of $\tau_Q\to\infty$, $S^z(0)$  approaches the 
equilibrium value, which in the  thermodynamically-large quantum Ising model at $g=1$ 
is equal to $2/\pi$. 
Our numerical simulations performed for $N\gg\hat\xi$, i.e. in the regime where the KZ theory 
is supposed to work, suggest that  
\be
S^z(0)-\frac{2}{\pi}\sim  \frac{1}{\sqrt{\tau_Q}},
\label{minus2pi}
\ee
which is discussed in Fig. \ref{szcrit_fig}. A similar result is obtained when
the  continuous  quench takes the Ising chain from one  phase to another
\cite{PuskarovSciP2016}.

\section{Dynamics of the transverse magnetization at the critical point}
\label{Dynamics_sec}
After arriving at the critical point, the system undergoes free evolution. 
The wave-function is again given by (\ref{wavef}). This time, however, 
the modes evolve according to 
\be
\ii \frac{d}{dt} 
\left ( \begin{array}{c}
v_k \\
u_k \\
\end{array}
\right )
= 
\tilde h_k
\left ( \begin{array}{c}
v_k \\
u_k \\
\end{array}
\right ), \
\tilde h_k = 
2 \left(\begin{array}{cc}
1 - \cos(k) & -\sin(k) \\
- \sin(k) &      \cos(k) - 1 \\
\end{array}\right).
\ee

\begin{figure}[t]
\includegraphics[width=\pref\columnwidth,clip=true]{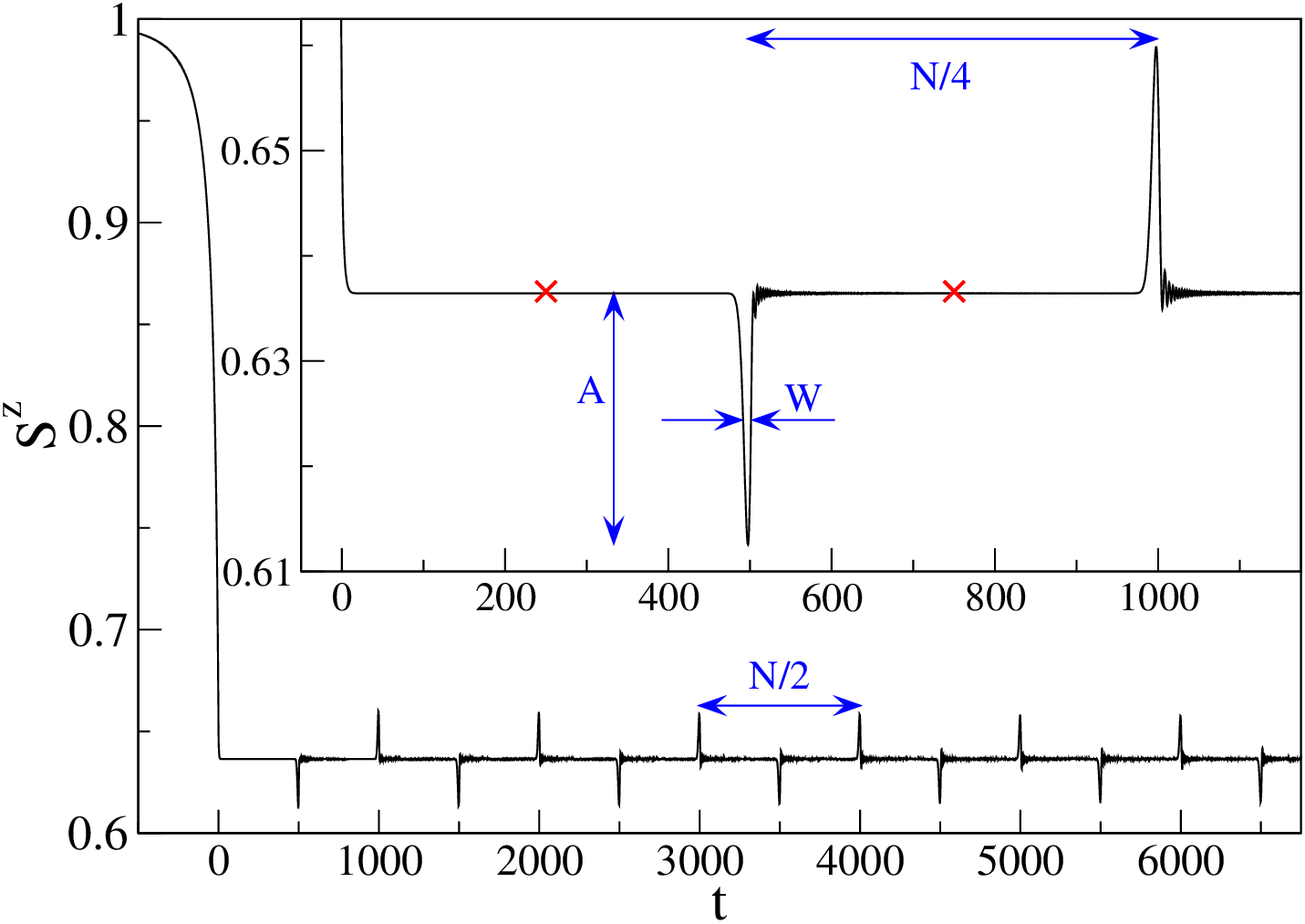}
\caption{Dynamics of the transverse  magnetization (\ref{Sz})  for quench protocol (\ref{qprotocol}). 
The inset zooms in the first two peaks that appear  in the course
of free  evolution. 
Note that we do not count the decay of magnetization  near $t=0$ as
 a peak. The amplitude and half-width of the first peak are marked as $A$ and $W$, respectively.
The red crosses show the
equilibrium value of the $S^z$ magnetization at the critical point 
of a thermodynamically-large system: $2/\pi\approx0.637$.
The system size $N=2000$ and the quench time $\tau_Q=100$.
}
\label{free_fig}
\end{figure}

Solving this equation with  initial conditions that $v_k(0)$ and $u_k(0)$
are computed from  (\ref{vkodt}) and (\ref{ukodt}), respectively, we get
\be 
\left( \begin{array}{c}
v_k(t) \\
u_k(t) \\
\end{array}\right) 
= 
\left(\begin{array}{cc}
\cos(\varepsilon_k t) - \ii \sin(k/2) \sin( \varepsilon_k t) & \ii \cos(k/2) \sin(\varepsilon_k t) \\
\ii \cos(k/2) \sin(\varepsilon_k t) & \cos(\varepsilon_k t) + \ii \sin(k/2) \sin(\varepsilon_k t) \\
\end{array}\right)
\left( \begin{array}{c}
v_k(0) \\
u_k(0) \\
\end{array}\right),
\label{free_odt}
\ee
where 
\be
\varepsilon_k=4\sin(k/2).
\label{ek}
\ee
Eigenvalues of the Hamiltonian $\tilde h_k$ are equal to $\pm\varepsilon_k$.

Having  $v_k(t)$ modes, we can compute the  transverse magnetization along the field direction (\ref{Sz})
\be
S^z(t)=1-\frac{4}{N}\sum_k\BBB{
[\cos(\varepsilon_k t) - \ii \sin(k/2) \sin( \varepsilon_k t)]v_k(0)+
\ii \cos(k/2) \sin(\varepsilon_k t) u_k(0)}^2.
\label{Szfull}
\ee
A typical result that we get is presented in Fig. \ref{free_fig}, where both 
driven evolution to the critical point and   free evolution at the
critical point are plotted. We will focus on the latter below.

\begin{figure}[t]
\includegraphics[width=\pref\columnwidth,clip=true]{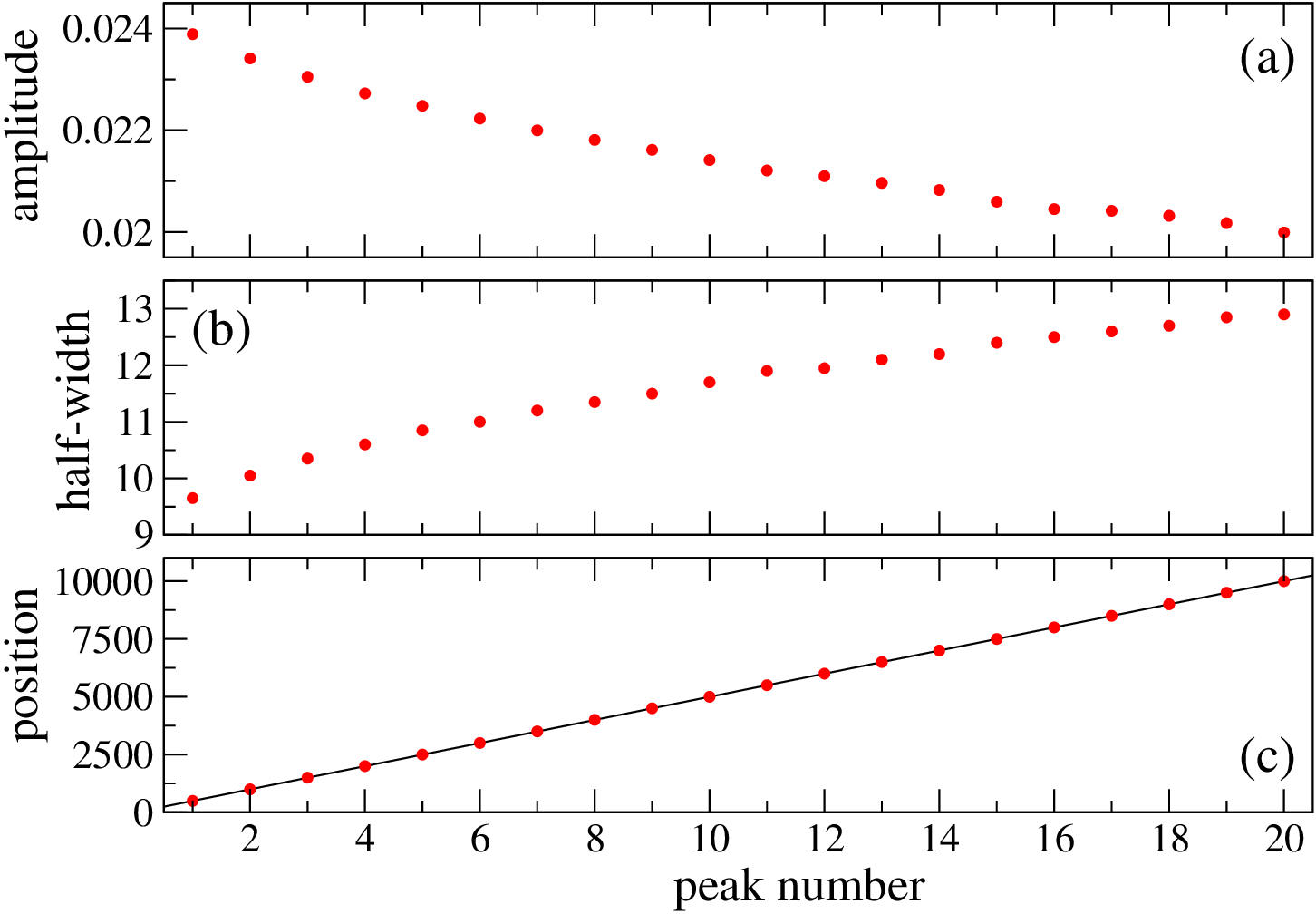}
\caption{Illustration of quasi-periodicity of  oscillations of
the transverse 
magnetization at the critical point. Panels (a) and (b) show  the  amplitude and half-width of  subsequent peaks--see 
Fig. \ref{free_fig}.
They show  that  peaks gradually become smaller in the amplitude and broader in the time domain.
Panel (c) shows position in the time domain of  subsequent peaks.  The
solid line is the linear fit $-2.66(4)+500.079(4)n$, where $n$ is the
peak number. The distance  between the peaks is nearly equal to $N/4=500$. It 
 does  not exhibit a gradual drift seen in panels (a) and (b).
The plots are done for $N=2000$ and $\tau_Q=100$, just as in Fig.
\ref{free_fig}. The dots show numerics.
}
\label{peaks_fig}
\end{figure}

The main  feature that we see in Fig. \ref{free_fig} is the 
series of quasi-periodic peaks in 
the transverse magnetization appearing during free evolution (we will call as peaks {\it both} minima and
maxima, such as those separated by $N/4$ in the inset of Fig. \ref{free_fig}). We use the 
term quasi-periodic to  indicate that the width and amplitude of the peaks is not constant. This 
is illustrated in Fig. \ref{peaks_fig},
where we see that the amplitude (width) of  subsequent peaks
is gradually decreasing (increasing).
Note that the peaks  appear at
nearly identical time intervals.

To explain quasi-periodicity from Fig. \ref{free_fig}, we note that after setting 
\be
\varepsilon_k\approx 2k
\label{ek_approx}
\ee
in (\ref{Szfull}), and using an expression for quasi-momenta (\ref{kk}),  
we get a periodic in time  solution for $S^z(t)$ with the repetition period equal to $N/2$. 
Such an approximation is valid for small-momenta only, i.e.,  $k\ll\pi$.
As a result, instead of a perfectly periodic pattern for $S^z(t)$, we get 
the quasi-periodic one. 
This can be understood in the following way. If we drove the
system perfectly adiabatically to the critical point, we would see  no
oscillations of magnetization during free evolution as the system would be in
the ground state. Oscillations that we see come from system
excitation, which mainly affects the modes satisfying condition (\ref{khat}). 
For such modes approximation (\ref{ek_approx}) is justified, leading to $N/2$
quasi-repetition period. This time scale is clearly associated with  
the characteristic time scale $\tau_c$ 
appearing at
the critical point (\ref{wczasy}).

Next, we would like to understand how the width and amplitude of the peaks 
depend on the quench rate. As the analytical evaluation of (\ref{Szfull}) is impossible,
we will again restore to an approximation. Namely, we will  put expressions
(\ref{approx1}) and (\ref{approx2}) into (\ref{Szfull}) and sum over all
momenta (\ref{kk}). Such a procedure  should reasonably  account for the
contribution of  excited  modes and ignore the contribution of the
ones that evolved nearly adiabatically. The latter, however, do not
contribute to the dynamics.
In the end, keeping only the terms in the leading order in $k\sqrt{\tau_Q}$,
we arrive at
\be
S^z(t)\approx\text{const} -\frac{4}{N}\sum_k\BB{\sin^2(\varepsilon_k t)
+\frac{\sqrt{\pi}}{2}k\sqrt{\tau_Q}\sin(2\varepsilon_k t)+
\frac{\pi}{2}k^2\tau_Q\cos^2(\varepsilon_k t)}e^{-\pi\tau_Qk^2/2},
\label{Szseriest}
\ee
where the constant term is time-independent. It does, however, depend on $\tau_Q$
due to (\ref{minus2pi}).

\begin{figure}[t]
\includegraphics[width=\pref\columnwidth,clip=true]{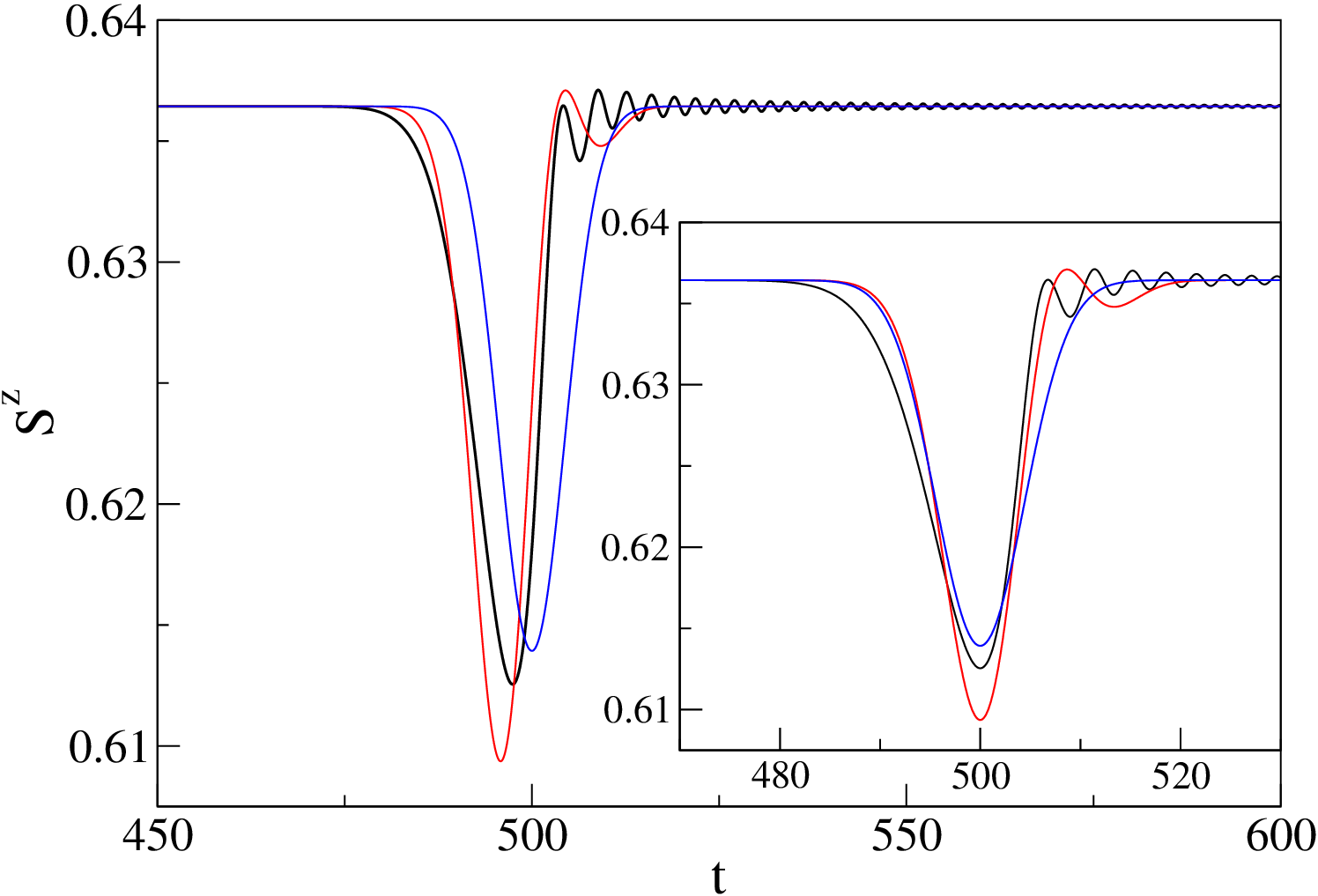}
\caption{The first peak in the  transverse magnetization during free time evolution.
The black line shows numerics and comes from Fig. \ref{free_fig}.
The red line shows  (\ref{Szseriest}), while the blue line shows
(\ref{main2}). The constant  terms in (\ref{Szseriest}) and (\ref{main2}) have
been chosen to reproduce  numerical results far away from the peak center.
The inset shows the same as
the main plot, but with data shifted horizontally so that the peaks are centered around the same time.
The plots are done for $N=2000$ and $\tau_Q=100$. 
}
\label{single_mag_fig}
\end{figure}

The leading contribution in this expression comes from the zeroth-order term
in $k\sqrt{\tau_Q}$, which we additionally simplify with (\ref{ek_approx}) and 
identity $\sin^2(2kt)=1/2-\cos(4kt)/2$
\be
S_0^z(t)=\text{const}+\frac{2}{N}\sum_k \cos(4kt)e^{-\pi\tau_Qk^2/2},
\label{main1}
\ee
where $-2/N\sum_k e^{-\pi\tau_Qk^2/2}$ has been absorbed into the constant
term.
By putting quasi-momenta (\ref{kk}) into (\ref{main1}), 
we immediately see that  its  time-dependent component is anti-periodic
with anti-period equal to $N/4$. 
This expression describes the peaks, whose width and shape can be obtained
by  replacement (\ref{X}),
which amounts to taking the thermodynamic limit.
Doing so, one finds that the right-hand-side of (\ref{main1}) can be
approximated for $\tau_Q\gg1$ by
\be
\text{const}+\frac{1}{\pi\sqrt{2\tau_Q}}e^{-8t^2/(\pi\tau_Q)}.
\ee
The time-dependent part of such a result, however, is not anti-periodic because in the
limit of $N\to\infty$ the anti-period of oscillations goes to infinity.
The missing anti-periodicity can be recovered by writing the solution  in the following form
\be
S_0^z(t)=\text{const}+\frac{1}{\pi\sqrt{2\tau_Q}}
\sum_{s=0}^\infty
(-)^{s}e^{-8(t-sN/4)^2/(\pi\tau_Q)}.
\label{main2}
\ee
The differences between (\ref{main1}) and (\ref{main2}) are negligible when
$N\gg1$ and the width of the peaks is small compared to the period of oscillations.
Approximations (\ref{Szseriest}) and (\ref{main2}) are compared to numerics around the first peak  in
Fig. \ref{single_mag_fig}.

Using (\ref{main2}), one easily gets that the amplitude of the peaks is 
\be
A=\frac{1}{\pi\sqrt{2\tau_Q}}\approx\frac{0.225}{\sqrt{\tau_Q}},
\label{Aana}
\ee
while their half-width is 
\be
W=\sqrt{\frac{\pi\ln2}{2}}\sqrt{\tau_Q}\approx1.04\sqrt{\tau_Q}.
\label{Wana}
\ee
Several remarks are in order now. 

\begin{figure}[t]
\includegraphics[width=\pref\columnwidth,clip=true]{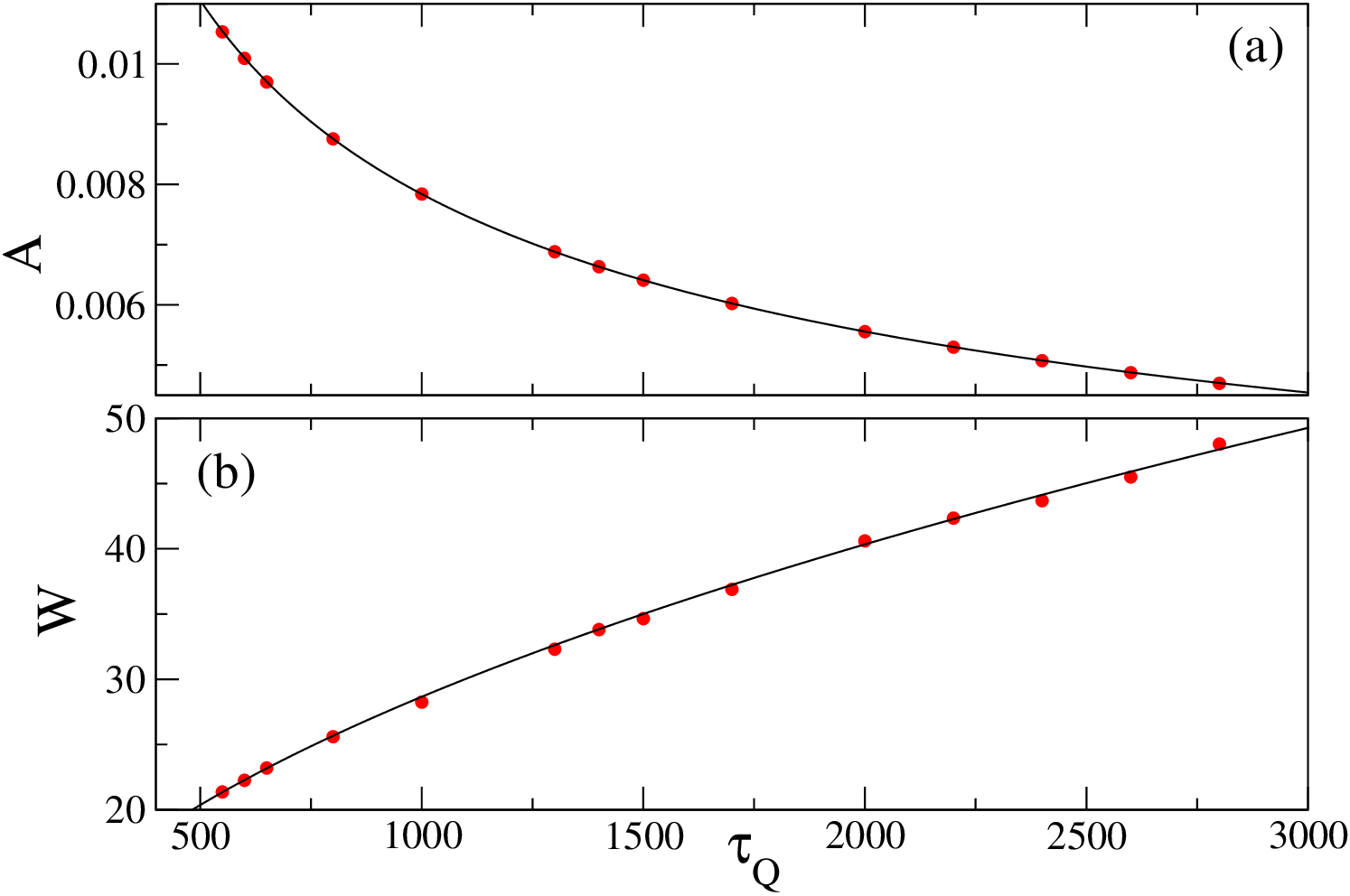}
\caption{Properties of the first peak in the transverse magnetization as a function of the quench
time $\tau_Q$.
Panel (a): the amplitude of the first peak. Red dots show numerics, while the  solid black line comes from the
fit $\ln(A)=-1.421(2) - 0.4962(3)\ln(\tau_Q)$. This can be
rewritten as $A\approx0.241/\tau_Q^{0.496}$ and compared to (\ref{Aana}). 
Panel (b): the half-width of the first peak. Again red dots show numerics, while the solid black line comes from the
fit $\ln(W)=-0.05(3) + 0.493(4)\ln(\tau_Q)$, which can be
reworked  to $W\approx0.951\tau_Q^{0.493}$ and compared to (\ref{Wana}).
The system size $N=2000$ in both panels.}
\label{AW_fig}
\end{figure}

First of all, it should be stressed  that (\ref{main2}) is derived under 
approximation (\ref{ek_approx}) and so its agreement with numerics should 
become gradually worse as time  increases. Indeed, as we know from Fig.
\ref{peaks_fig}, the amplitude and half-width of the peaks varies in time in
contrast to what (\ref{Aana}) and (\ref{Wana}) predict.
Therefore, we will focus below on the discussion of how these two analytical predictions 
compare to numerics around the first peak, where our approximations  should work best. 
Such numerics is discussed in Fig. \ref{AW_fig}. 
The $\tau_Q$  scaling exponents in (\ref{Aana}) and (\ref{Wana}) 
match it  within about $1\%$, which is a very good result. 
The prefactors  in (\ref{Aana}) and (\ref{Wana}) differ by about $8\%$ from the ones obtained numerically for the first peak,
which is  reasonable given the low order of the approximation.

Second, half-width  (\ref{Wana}) scales with $\tau_Q$ in the same way
as  $\hat t$ (see also discussion by the end of Sec. \ref{Loschmidt_sec}).
Thus, we see in the non-equilibrium dynamics at the critical point two
characteristic time scales identified  in Sec. \ref{Introduction}, $\hat t$ and $\tau_c$, 
that are given in the quantum Ising model by (\ref{wczasy}). 

Third, we get by combining (\ref{Aana}) and (\ref{Wana}) 
\be
A\cdot W \approx 0.23,
\ee
which, quite interestingly, is $\tau_Q$-independent in the quantum Ising model. Nearly identical result
can be obtained from  the fits to numerical data discussed in Fig. \ref{AW_fig}.

Finally, we note that when the quench does not end at the critical point,
different  dynamics of the transverse magnetization during subsequent free evolution is observed 
\cite{PuskarovSciP2016}.

\section{The Loschmidt echo at the  critical point}
\label{Loschmidt_sec}
Using the  results  from Sec. \ref{Dynamics_sec}, we can analyze the Loschmidt
echo,
which is defined here as the squared  overlap between the
initial state, say $|\phi\ra$, and the time-evolved state, $e^{-\ii\hat {\cal H} t}|\phi\ra$.
Such a quantity--also known as the return amplitude or  fidelity--evolves 
in time whenever  $|\phi\ra$ is not an
eigenstate of the Hamiltonian $\hat {\cal H}$.
It has been recently intensively analyzed for sudden quenches, where $|\phi\ra$ is 
an eigenstate of the pre-quench Hamiltonian, while $\hat{\cal H}$ is the post-quench  Hamiltonian 
(see  e.g. \cite{HeylPRL2013,CardyPRL2014,NajafiPRB2017} for studies relevant to the 
quantum Ising model and \cite{ZvyaginLow2016,HeylArxiv2017} for reviews). 

We are interested in the Loschmidt echo at the critical point, $\hat{\cal H}=\hat H(g=1)$, computed  
for  $|\phi\ra$ being the  non-equilibrium state of the quantum Ising model  driven to the
critical point
\be
L(t)=|\la\psi(0)|e^{-\ii\hat H(g=1)t}|\psi(0)\ra|^2=\BBB{\la\psi(0)|\psi(t)\ra}^2,
\ee
where   $|\psi(t)\ra$  is obtained by combining (\ref{wavef}) with
(\ref{free_odt}). A similar strategy for  studies of the Loschmidt echo in the
quantum Ising model was 
employed in \cite{PuskarovSciP2016}, where a continuous quench starting and ending away from the critical point 
was used to prepare the state $|\phi\ra$. This led to the dynamics of  the Loschmidt echo,
which differs from what  we report below. We mention in passing that studies  in  \cite{PuskarovSciP2016} 
revolve around the  aspects  of the Loschmidt echo that are different from the
ones we focus on.

\begin{figure}[t]
\includegraphics[width=\pref\columnwidth,clip=true]{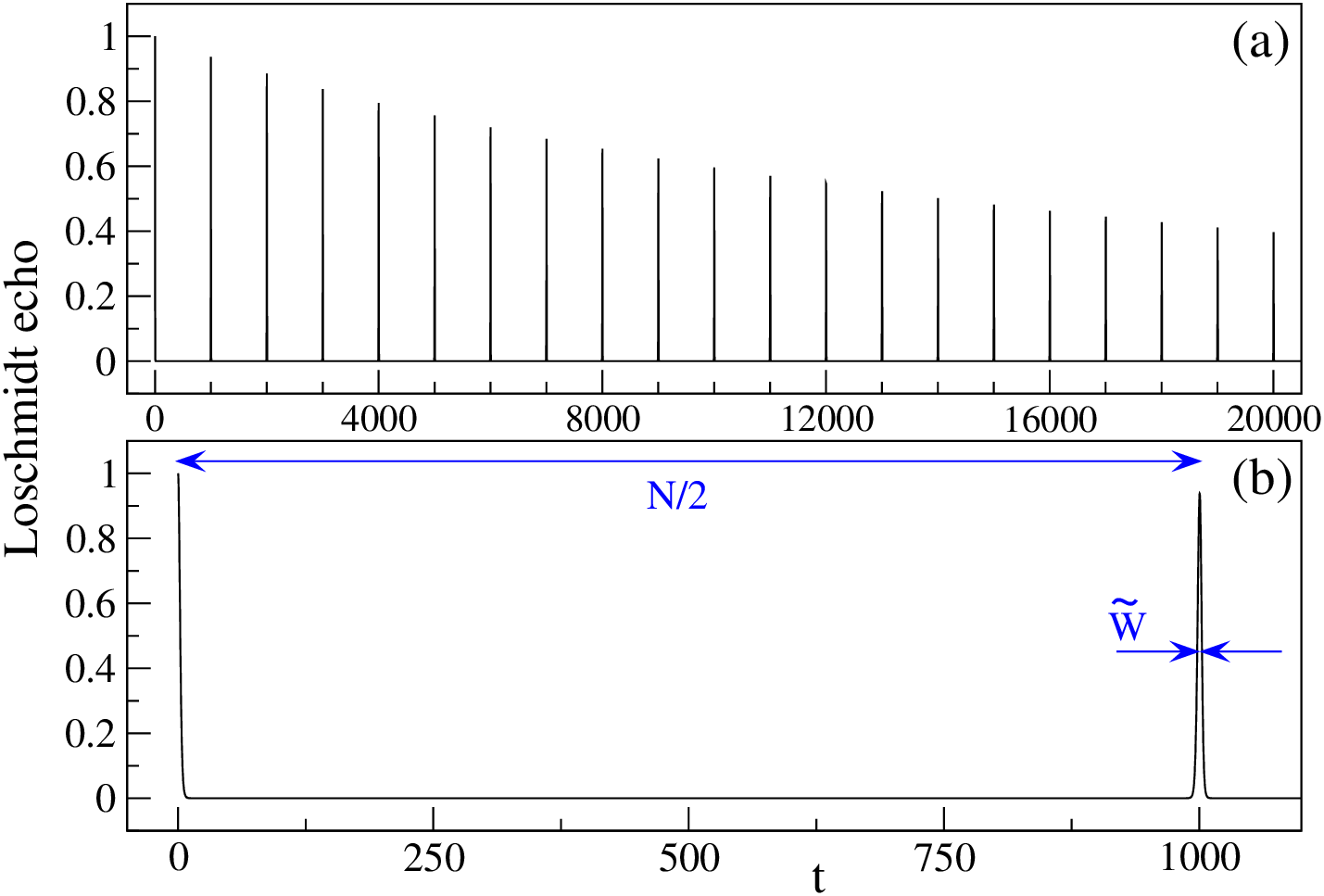}
\caption{The Loschmidt echo for $N=2000$ spins and the quench time $\tau_Q=100$.
Panel (a) illustrates periodicity and decay of echo revivals. Panel (b) 
shows the Loschmidt echo for small times up to the first peak. 
Note that we do not count the decay of the Loschmidt echo near $t=0$ as
a  peak. The half-width of the first peak is marked as $\tilde W$.
}
\label{loschmidt_fig}
\end{figure}

\begin{figure}[t]
\includegraphics[width=\pref\columnwidth,clip=true]{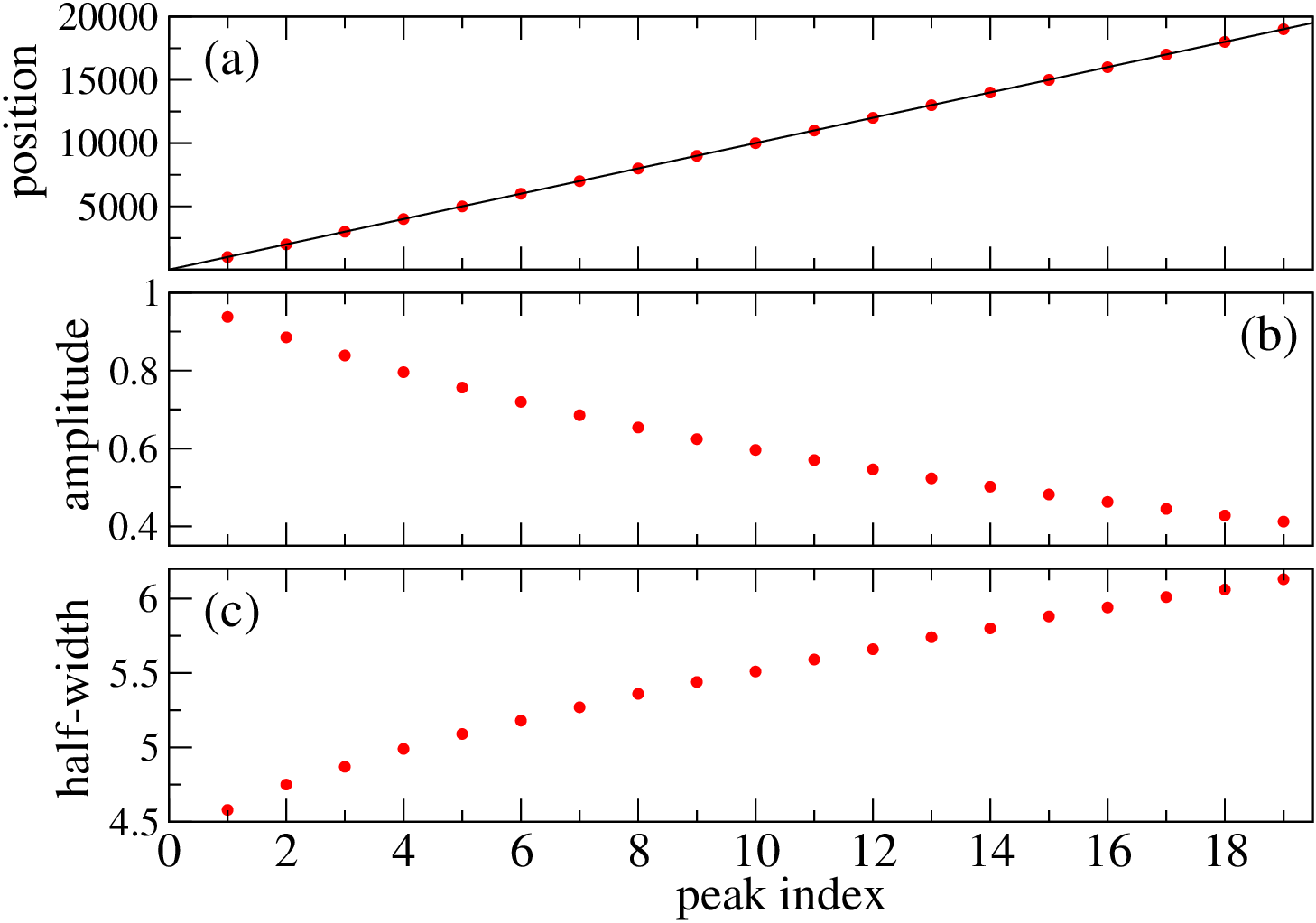}
\caption{Illustration of quasi-revivals of the Loschmidt echo.
Panel (a) shows position in the time domain of the subsequent peaks.  The
solid line is the linear fit $0.96(23)+1000.11(2)n$, where $n$ is the
peak number. The distance  between the peaks is nearly equal to $N/2=1000$. It 
 does  not exhibit a gradual drift with the peak number seen in panels (b) and (c) showing 
 the  amplitude and half-width of the subsequent peaks.
These two panels  show  that the peaks gradually become smaller in amplitude and broader in the time domain.
The plots are done for $N=2000$ and $\tau_Q=100$. The dots come from numerical
simulations.
}
\label{ov_peaks_fig}
\end{figure}

Typical results that we obtain are presented in Fig. \ref{loschmidt_fig}.
Just as in Sec. \ref{Dynamics_sec}, we observe peaks appearing at virtually identical time intervals 
(Fig. \ref{ov_peaks_fig}a), 
which become shorter and broader 
in the course of time evolution (Figs. \ref{ov_peaks_fig}b and \ref{ov_peaks_fig}c). 
The  Loschmidt echo is nearly zero between
the peaks. 
Analytical insights into some of the properties of the Loschmidt echo can be worked
out in the following way.

We use  (\ref{wavef}) and the normalization condition 
$|u_k(t)|^2+|v_k(t)|^2=1$ to find 
\be
\label{Lnext1}
L(t)=\prod_k \B{1-|u_k(0)v_k(t) - v_k(0)u_k(t)|^2}.
\ee
Such  an expression for  $L(t)$ allows us to use approximations (\ref{approx1}) and
(\ref{approx2}) for arbitrary $k$. Indeed, in the limit of small $k$, in the sense of
(\ref{khat}), such approximations faithfully represent the exact result. In the
opposite limit,  due to the Gaussian cutoff (\ref{Gcutoff}) in (\ref{approx1})
and (\ref{approx2}), we are left in
 (\ref{Lnext1}) with the product of unities representing the fact that
 large-enough $k$ Bogolubov modes do not get excited during  time evolution
towards the critical point. Such modes  do not contribute to the Loschmidt echo.
We note in passing that 
we compute 
the probability of finding the Ising chain in the
ground state at $t=0$ through the  product of $1-p_k$ in   (\ref{pgs}) due to exactly the same reasons.

\begin{figure}[t]
\includegraphics[width=\pref\columnwidth,clip=true]{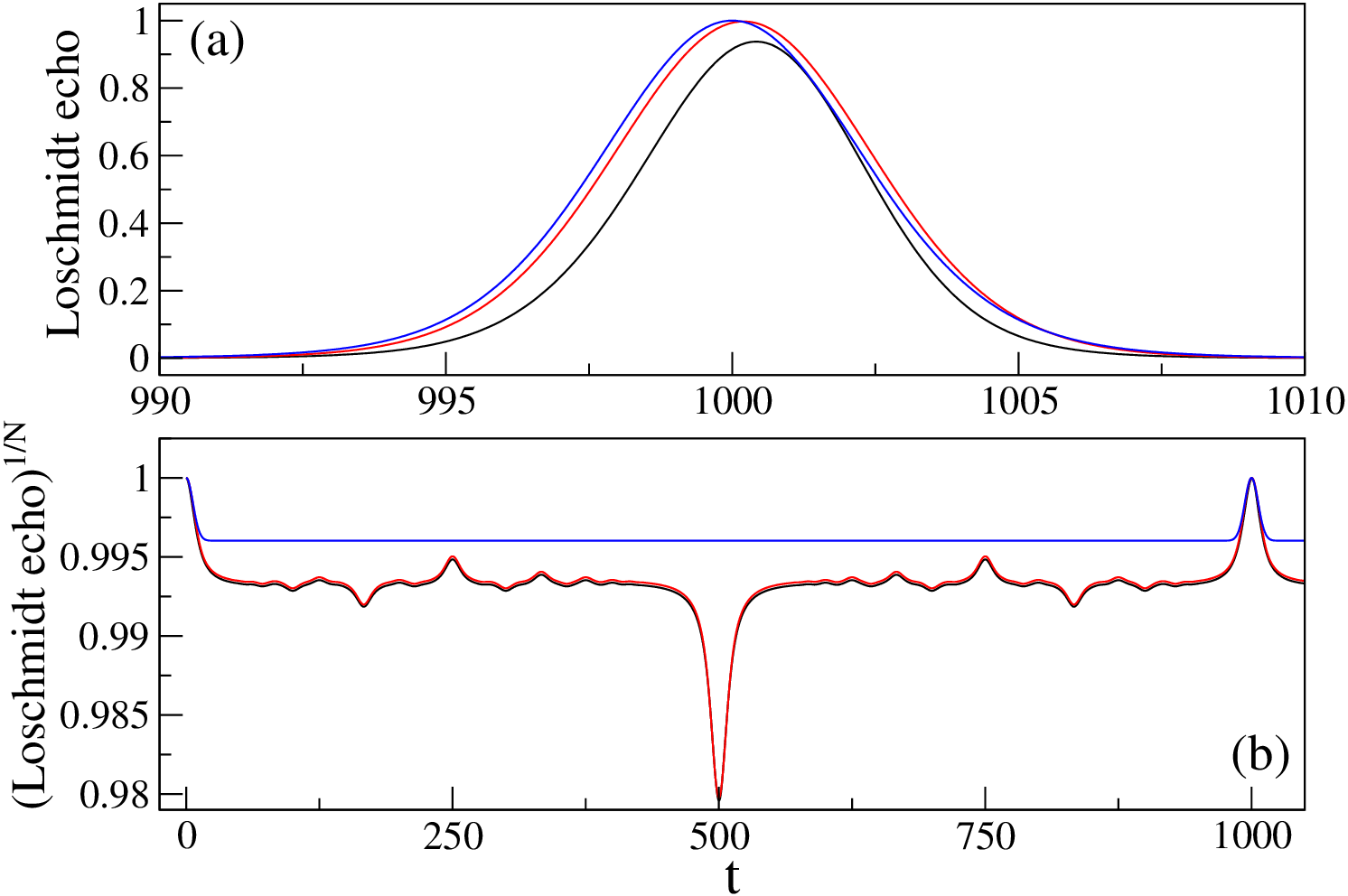}
\caption{Panel (a) shows comparison between the exact numerical results for
the Loschmidt echo near the first peak (black
line) and approximations  (\ref{Lprodek}) and  (\ref{Lthermo}) given by the
red and blue lines, respectively. 
Panel (b) shows data from panel (a) taken to the
power of $1/N$ and displayed in a wider time range. 
In this way the Loschmidt echo between the peaks can be efficiently
shown  so that quality of the analytical approximations between the peaks
can be critically assessed.
The plots are done for $N=2000$ and $\tau_Q=100$. 
}
\label{single_ov_fig}
\end{figure}

Proceeding similarly as in Sec. \ref{Dynamics_sec}, we arrive after expansion
in $k\sqrt{\tau_Q}$ at
\be
L(t)\approx\prod_k\BB{1-\sin^2(\varepsilon_k t)e^{-\pi\tau_Qk^2}},
\label{Lprodek}
\ee
which can be compared to (\ref{Szseriest})--note that  terms linear and
quadratic in $k\sqrt{\tau_Q}$ vanish here. This can be further simplified by
using (\ref{ek_approx}) to get 
\be
L_0(t)=\prod_k\BB{1-\sin^2(2k t)e^{-\pi\tau_Qk^2}},
\label{poikk}
\ee
which is periodic in time with period $N/2$.
Next,  in the spirit of the discussion from the previous section, we take the
thermodynamic limit (\ref{X}) on the right-hand-side of (\ref{poikk})  to obtain
\begin{align}
\exp\B{\frac{N}{2\pi}\int_0^\pi dk \ln\BB{1-\sin^2(2kt)e^{-\pi\tau_Qk^2}}}.
\label{derf}
\end{align}
This can be further simplified by noting that the argument of the exponent is peaked
around $t=0$, where it can be approximated by
\be
-\frac{N}{2\pi}\int_0^\pi dk \sin^2(2kt)e^{-\pi\tau_Qk^2}\approx
-\frac{N}{8\pi\sqrt{\tau_Q}}\B{1-e^{-4t^2/(\pi\tau_Q)}}
\label{llll}
\ee
for   $\tau_Q\gg1$.
Restoring ``by hand'' $N/2$ periodicity, which we have lost by taking the thermodynamic limit in (\ref{derf}),  we arrive at 
\be
L_0(t)\approx\exp\BB{-\frac{N}{8\pi\sqrt{\tau_Q}}\B{1-\sum_{s=0}^\infty
e^{-4(t-sN/2)^2/(\pi\tau_Q)}}}.
\label{Lthermo}
\ee
Several remarks are in order now.

\begin{figure}[t]
\includegraphics[width=\pref\columnwidth,clip=true]{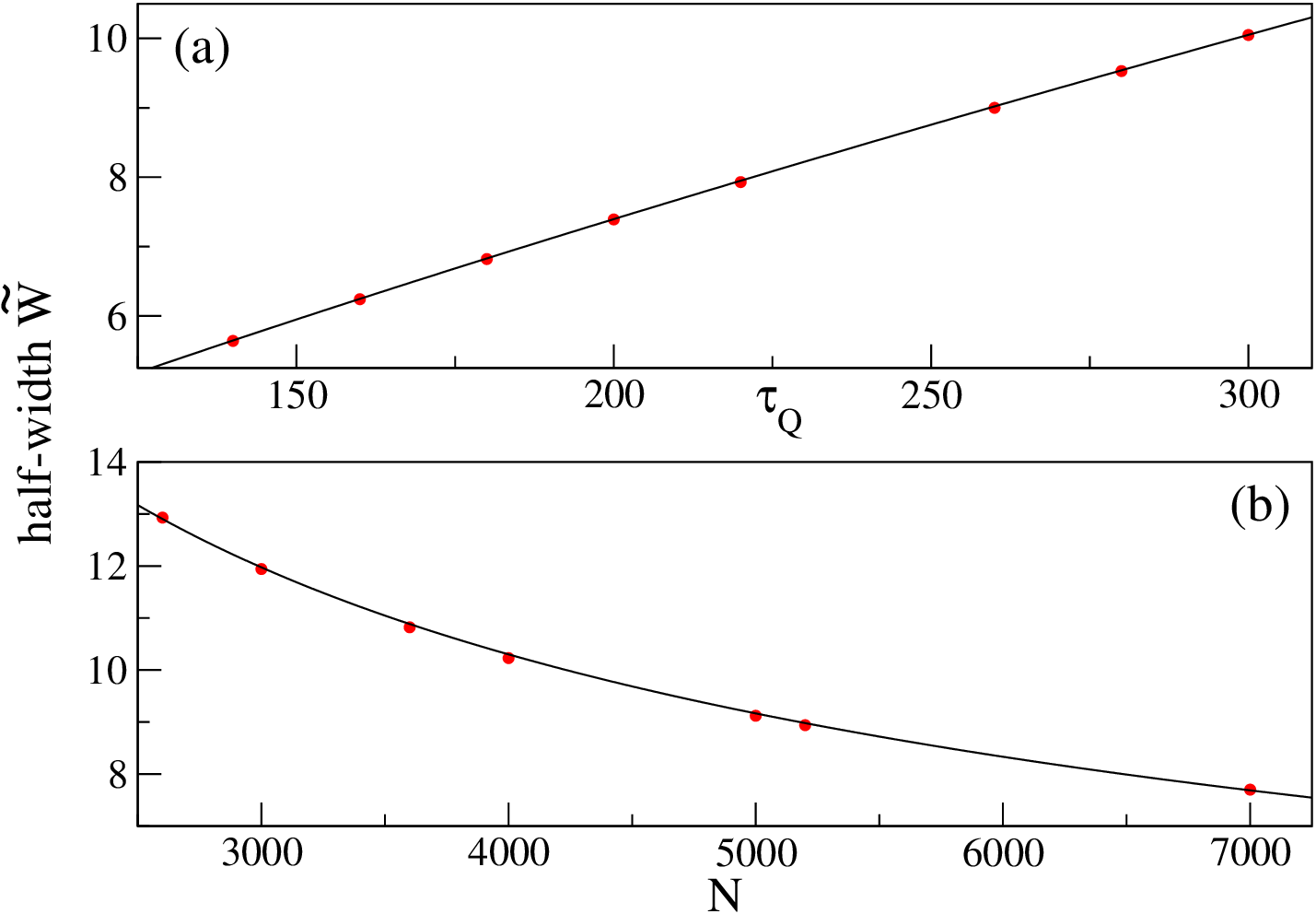}
\caption{
Panel (a): the half-width of the first peak of the Loschmidt echo for the system size $N=2000$. 
Solid black line comes from the fit $\ln(\tilde W)=-2.01(1)+0.757(1)\ln(\tau_Q)$, 
which can be rewritten as $\tilde W\approx 0.13\tau_Q^{0.757}$. This fit can
be  compared
to (\ref{tildeW}) evaluated at $N=2000$, i.e., to $0.165\tau_Q^{3/4}$.
Panel (b): the half-width of the first peak of the  Loschmidt echo for the
quench time $\tau_Q=500$. 
Solid black line comes from the fit 
$\ln(\tilde W)=6.67(4)-0.523(5)\ln(N)$, which can be recast to $\tilde
W\approx788.4/N^{0.523}$. That fit can be compared to (\ref{tildeW}) after setting
$\tau_Q=500$, which results in $\tilde W\approx782.2/N^{1/2}$. 
Red dots depict numerics in both panels.  
}
\label{WW_fig}
\end{figure}

First,  we would like to compare (\ref{Lprodek}) and (\ref{Lthermo}) to
numerics for small times, where they should work best. 
Fig. \ref{single_ov_fig}a shows that  around the maximum of the first peak
both approximations provide similar results in reasonable agreement with
numerics. Between the peaks, however, 
(\ref{Lprodek}) significantly outperforms (\ref{Lthermo}), which is shown in
Fig. \ref{single_ov_fig}b. There is, in fact, quite a good agreement between 
(\ref{Lprodek}) and numerics there.
Thus, the dispersion relation (\ref{ek}) is the key to 
getting the numerous features of the Loschmidt echo between the peaks (similar
features are seen in Fig. 2 of \cite{CardyPRL2014}, where the Loschmidt echo
for the  sudden quench to the critical point is discussed). 

Second, (\ref{Lthermo}) predicts perfect revivals of the Loschmidt echo at 
time intervals equal to $N/2$. While such periodicity is clearly seen in the
numerical data (Fig. \ref{ov_peaks_fig}a),  the revivals  that we
observe are only partial  (Fig. \ref{ov_peaks_fig}b). 

Third, the half-width $\tilde W$ of the peaks of (\ref{Lthermo})  is 
\be
\tilde W\approx\sqrt{-\pi\tau_Q\ln\BBB{1-\frac{8\pi\sqrt{\tau_Q}\ln2}{N}}}\approx
2\pi\sqrt{2\ln2}  \frac{\tau_Q^{3/4}}{\sqrt{N}}\approx7.4
\frac{\tau_Q^{3/4}}{\sqrt{N}},
\label{tildeW}
\ee
where condition (\ref{Nggsqrt}) has been employed to simplify the logarithm. This result does not predict
an increase of the
half-width with peak index, which we  see in Fig. \ref{ov_peaks_fig}c. 
Therefore,  we
will focus our attention on the half-width of the first peak, which we numerically study in Fig. \ref{WW_fig}. 
We see from the fits in this figure that 
the scaling exponents of (\ref{tildeW})  agree with numerics 
within a few percent, which is a very good result. The prefactor in (\ref{tildeW}) agrees with numerics to a lesser degree,
i.e., within
about $20\%$. 

Dependence of $\tilde W$ on the quench time $\tau_Q$ and the system size $N$ is quite unusual at first glance. 
Working backwards through the calculation and restoring physical dimensions,  we find that the right-hand-side of 
(\ref{tildeW}) can be written as
\be
\tilde W \sim \hat t \sqrt{\frac{\hat\xi}{L}},
\label{tilWscal}
\ee
where $L$ is the system size equal to $N$ times the lattice constant.
$L$ appears in (\ref{tilWscal}) from changing the sum over momenta into
the integral in (\ref{derf}). $\hat\xi$ comes from the Kibble-Zurek-like  assumption that universal 
dynamics of the system  depends on momenta through the $k/\hat k\sim k\hat\xi$ combination.
Rescaling momenta under  integral (\ref{llll})
\be
k\to k/\hat\xi
\label{Y}
\ee
produces the $\hat\xi$ factor in
(\ref{tilWscal}). $\hat t$
enters (\ref{tilWscal}) through  the dispersion relation. To show this, we note that
time enters our equations through $\varepsilon_k t\sim k^z t$ combination for
low-energy modes at the critical point, whose excitation produces revivals in the Loschmidt echo. 
Then, we use rescaling  (\ref{Y}) so that 
$\varepsilon_k t\to k^z t/\hat\xi^z$, and note that $\hat\xi^z\sim\hat t$ in
the Kibble-Zurek theory.
This produces $t/\hat t$ combination setting  the time scale for (\ref{tilWscal}).

Finally, we mention that similar reasoning can be used to argue that 
the half-width $W$ and amplitude $A$ of the peaks of transverse magnetization should   scale in the 
quantum Ising model as $\hat t$ and $1/\hat\xi$, respectively. This
observation agrees with the results reported in Sec. \ref{Dynamics_sec}.

\section{Summary}
\label{Summary_sec}
We have considered  a continuous quench  of the
quantum Ising chain that was initially prepared in the ground state far away from the critical point 
in the paramagnetic phase. 
The quench has brought the chain to the critical point, where we have  analytically determined 
how the probability of finding our system   in the ground state depends on the quench rate. 
We have  verified this result numerically and explained how it follows from
the Kibble-Zurek theory of non-equilibrium phase transitions. 
In the context of the Kibble-Zurek theory, previous studies of the probability of 
finding the system in the ground state have been  focused on quenches that ended 
far away from the critical point \cite{JacekPRB2009,MarekBodzioPRA,BDproceedings}. 
Thus, our results provide complementary insights into the process of excitation
during non-equilibrium  phase transitions.

Next, we have  focused our attention on the free evolution of such an out of equilibrium   system
at  the critical point. 
We have numerically studied   dynamics of its  
transverse magnetization and Loschmidt echo. 
We have found a series of quasi-periodic peaks in both observables  and  proposed  a simple analytical
model predicting some of their properties. In particular, such  a model
shows    that the  peaks should appear at time intervals proportional to  the
system size and that their width should scale with the quench rate in a way 
consistent with the  Kibble-Zurek theory.
These  predictions  have  been compared to numerics 
for small times of free evolution 
and good overall agreement have been found.

It is perhaps worth to stress that our studies of dynamics of
transverse magnetization did not employ the thermodynamic-limit  approximation routinely used in
the  earlier works, see e.g.  \cite{Barouch1970, PuskarovSciP2016}. This is important for
two reasons. First, it allowed us to observe  quasi-periodic peaks whose
repetition period diverges in the thermodynamic limit. Second, finite-size
results should be of experimental relevance in cold atom and ion quantum
simulators, whose size is nowhere near the thermodynamic limit
(see e.g. \cite{LukinNature2017,MonroeNature2017} for fascinating recent 
experimental progress in studies of dynamics of quantum phase transitions).

We would also like to point that our studies of the Loschmidt echo at the
critical point, unlike
some former works   \cite{CardyPRL2014,NajafiPRB2017},
involved a rather non-trivial initial state. Such a state was generated by the continuous quench
imprinting  universal critical exponents $z$ and $\nu$ onto the 
wave-function \cite{JacekAdv2010,Dutta2015,PolkovnikovRMP2011}. 
This made the Loschmidt echo a richer object to study. Indeed, not only
quantum revivals as in \cite{CardyPRL2014,NajafiPRB2017} can be studied 
in such a system, but also the details of the universal non-equilibrium
dynamics.

Coming back to the discussion from Sec. \ref{Introduction}, we note that
post-quench oscillations of magnetization at the critical point in the  spin-1 Bose-Einstein 
condensate \cite{BDNJP2008} and the  quantum Ising model    are
quite different. Indeed, the period of such oscillations in the latter system is 
not only $\hat t$-independent but also   divergent in
the thermodynamic limit. Further studies are needed for explanation of this key  difference.

Finally, we would like to mention two possible extensions of these studies.  
First, we believe that it would be interesting to find out if similar  non-equilibrium dynamics 
can be observed in other integrable  systems characterized by different critical exponents.
Our preliminary results obtained for the Ising model with three-spin
interactions \cite{WolfPRL2006}, where $z=2$ and $\nu=1$, show that magnetization
rapidly oscillates during post-quench dynamics at the critical point instead
of exhibiting distinct quasi-periodic peaks such as those illustrated in Fig.
\ref{free_fig}. 
Second, it would be also interesting to investigate the corresponding dynamics in near-integrable and
non-integrable systems such as the Ising model in the transverse and longitudinal
fields \cite{Coldea}, the Bose-Hubbard model \cite{Lewenstein,KrutitskyPhysRep2015}, 
the  Dicke model \cite{GarrawayRoyal2011}, etc.  As all these
experimentally-accessible systems
undergo a quantum phase transition,  their  Kibble-Zurek
dynamics at the critical point can be studied. Such  dynamics, however, is expected to be considerably more
complicated than that of  integrable models (see \cite{LangenJSTAT2016} for
a recent review on the relaxation  dynamics of near-integrable systems).

\section*{Acknowledgments}
We thank Marek Rams for useful discussions.
MB and BD were supported by the Polish National Science Centre (NCN) grant
DEC-2016/23/B/ST3/01152.


\end{document}